\documentclass{article}
\usepackage{emulateapj}
\usepackage{apjfonts}

\begin{document}

\submitted{To appear in The Astrophysical Journal Letters}

\title{Age Constraints for an M31 Globular Cluster
from Main Sequence Photometry$^1$}

\author{Thomas M. Brown$^2$, Henry C. Ferguson$^2$, Ed Smith$^2$, 
Randy A. Kimble$^3$, 
Allen V. Sweigart$^3$, Alvio Renzini$^4$, R. Michael Rich$^5$, 
\& Don A. VandenBerg$^6$}

\begin{abstract}

We present a color-magnitude diagram (CMD) of the globular cluster
SKHB-312 in the Andromeda galaxy (M31), obtained with the Advanced
Camera for Surveys on the Hubble Space Telescope.  The cluster was
included in deep observations taken to measure the star formation
history of the M31 halo.  Overcoming a very crowded field, our
photometry of SKHB-312 reaches $m_V \approx 30.5$~mag, more than 1~mag
below the main sequence turnoff.  These are the first observations to
allow a direct age estimate from the turnoff in an old M31 cluster.
We analyze its CMD and luminosity function using a finely-spaced grid
of isochrones that have been calibrated using observations of Galactic
clusters taken with the same camera and filters.  The luminosity
difference between the subgiant and horizontal branches is
$\approx$0.2~mag smaller in SKHB-312 than in the Galactic clusters
47~Tuc and NGC~5927, implying SKHB-312 is 2--3~Gyr younger.  A
quantitative comparison to isochrones yields an age of
$10^{+2.5}_{-1}$~Gyr.

\end{abstract}


\section{Introduction}

For more than 50 years (e.g., Sandage 1953), main sequence photometry
has been the most direct method for determining the ages of both star
clusters and populations with more complex star formation histories.
Current age estimates for globular clusters in the Galactic halo
generally run from 11--13.5~Gyr (e.g., VandenBerg 2000).  Although
main sequence photometry has enabled the dating of globular clusters
in the satellites of the Milky Way (e.g., Rich et al.\ 2001), no
globular clusters have been imaged down to the main sequence in
another spiral galaxy.  Instead, extragalactic globular cluster ages
are inferred from composite colors and/or spectroscopy.  Recent
evidence suggests that Andromeda (M31) may have more young globular
clusters than the Galaxy, implying that M31 may have obtained more of
its clusters from cannibalized dwarf galaxies (Barmby \& Huchra 2000;
Burstein et al.\ 2004).

We obtained extremely deep images of M31 with the Advanced Camera for
Surveys (ACS) on the Hubble Space Telescope (HST) in order to measure
the star formation history in its halo (Brown et al.\ 2003).  Our
field, 51$^\prime$ from the nucleus on the southeast minor axis, was
chosen to include the M31 globular cluster SKHB-312 (Sargent et al.\
1977).  Spectroscopy (Huchra et al.\ 1991) and red giant branch (RGB)
photometry (Holland et al.\ 1997) yield respective metallicity
estimates of [Fe/H]~$=-0.7$ and [Fe/H]~$=-0.53$ for SKHB-312,
\linebreak

{\small \noindent $^1$Based on observations made with the NASA/ESA
Hubble Space Telescope, obtained at the Space Telescope Science Institute, 
which is operated by AURA, Inc., under NASA contract NAS 5-26555. These
observations are associated with proposal 9453.

\noindent $^2$Space Telescope Science Institute, 3700 San Martin Drive,
Baltimore, MD 21218;  tbrown@stsci.edu, ferguson@stsci.edu,
edsmith@stsci.edu 

\noindent $^3$Code 681, NASA Goddard Space Flight Center, Greenbelt, MD 20771; 
randy.a.kimble@nasa.gov, allen.v.sweigart@nasa.gov 

\noindent $^4$European Southern Observatory, Karl-Schwarzschild-Strasse 2, 
Garching bei M$\ddot{\rm u}$nchen, Germany; arenzini@eso.org 

\noindent $^5$Division of Astronomy, Dpt.\ of Physics \& Astronomy, UCLA, Los
Angeles, CA 90095; rmr@astro.ucla.edu

\noindent $^6$Department of Physics \& Astronomy, 
University of Victoria, P.O. Box
3055, Victoria, BC, V8W 3P6, Canada; davb@uvvm.uvic.ca
}

\noindent
near the peak in the metallicity distribution of
the M31 halo ([Fe/H]=$~-0.8$; Durrell et al.\ 2001).
Our cluster photometry reaches the main sequence in the F606W (broad
$V$) and F814W ($I$) bands, allowing a direct estimate of
its age.  Our analysis is based upon comparisons to photometry of
Galactic globular clusters in the same ACS bands,
and upon a finely-spaced grid of isochrones calibrated
using those clusters.

\section{Observations and Data Reduction}

These ACS observations are part of a large program to image the M31
halo, with 3.5 days of exposure time split between the F606W and F814W
filters; see Brown et al.\ (2003) for a description of the data and
its reduction.  The resulting color-magnitude diagram (CMD) extends
more than 1.5 mag below the old main sequence turnoff in the M31 halo.
We included the globular cluster SKHB-312 in our field so that the
turnoff could also be reached in an M31 cluster (Figure 1$a$).  Using
images from the Wide Field Planetary Camera 2 (WFPC2), Holland et al.\
(1997) determined the SKHB-312 tidal radius to be $\approx 10\arcsec$.
Although small compared to the ACS field ($\approx 205\arcsec \times
205\arcsec$), we placed the cluster near the field edge to minimize
the field area contaminated by the cluster in case it appeared more
extended in deeper observations. As expected for its high metallicity,
SKHB-312 has no known RR Lyrae stars, but it does have 13 long period
variables near the RGB tip (Brown et al.\ 2004); two are within the
area we ultimately analyzed.

We measured a more tractable $81\arcsec \times 48\arcsec$ subsection
around SKHB-312.  After constructing a variable point spread function
(PSF) from the most isolated stars in the cluster vicinity, we
performed aperture and PSF-fitting photometry using the DAOPHOT-II
package (Stetson 1987), with zeropoints matching the full field
catalog (Brown et al.\ 2003).  Our photometry is in the STMAG system:
$m= -2.5 \times $~log$_{10} f_\lambda -21.1$~mag.

We constructed our SKHB-312 CMD from stars within an an annulus chosen
to maximize the signal-to-noise ratio and minimize field
contamination.  The inner radius (100 pix; 3$\arcsec$) was where the
completeness dropped precipitously for stars near the turnoff (Figure
1$b$); the outer radius (300 pix; $9\arcsec$) was where the density
dropped to twice that in the field (Figure 1$c$). Because SKHB-312 was
near the field edge and the observations were dithered, the exposure
time was not uniform across the annulus.  We discarded the fraction of
the annulus ($< 0.5$\%) that had half of the total exposure time, but
kept the fraction (14\%) that was exposed for 75\%; 86\% of the
annulus was exposed fully.

We used extensive artificial star tests to determine the photometric
scatter and completeness as a function of color, luminosity, and field
position.  Small numbers of stars were repeatedly added to the images
with a distribution tracking the cluster light profile (to match these
tests to the data), with the noise reflecting the exposure time
variations.  Only those results within the 100--300 pix annulus were
used in our modeling.  We then retained those 1720 stars in the
SKHB-312 catalog above the 80\% completeness magnitude, with the
cutoff being a function of position (see Figure 1$b$), thus producing
a much cleaner CMD.  For consistency, we treated the cluster and field
contamination models the same way (\S3).

Our program included ACS observations of five Galactic clusters
spanning a wide [Fe/H] range, to provide cluster fiducials and to
verify the transformation of our isochrones to the ACS bands.  The
most relevant here are 47~Tuc and NGC5927, bracketing the SKHB-312
metallicity.  Our adopted parameters are listed in Table 1; these
values give consistency across the Galactic cluster observations and
imply one simple transformation of all isochrones to the ACS bands.
Note that our isochrones do not include core He diffusion, which would
reduce their ages by 10--12\% (VandenBerg et al.\ 2002).  After
reevaluating the range of parameters in the literature, the Galactic
cluster parameters have been updated slightly from those assumed in
our earlier papers (Brown 2003; Brown et al.\ 2003), and the changes
have been included in our calibration of the isochrone transformation.
The parameters for NGC5927 are highly uncertain, due to its high,
spatially variable reddening (Heitsch \& Richtler 1999), but the
values here are very close to those in Harris (1996).  We observed the
Galactic clusters with staggered exposure times to increase the
dynamic range, but the upper RGB saturated in the shortest possible
exposure of 0.5~s.

\section{Analysis}

We show the CMD of SKHB-312 in Figure 1$d$.  This is the first CMD to
resolve the main sequence in a globular cluster of M31.  The
horizontal branch (HB) consists of a tight red clump (unobscured in
Figure 1$e$), with no extension to a blue HB, while the RGB is very
red; these features are distinct from the CMD of the general field
population (Brown et al.\ 2003), and indicate that the cluster is more
metal rich than many of the field stars.  In this panel, we also plot
the ridge lines ({\it curves}) and HB loci ({\it points}) of 47 Tuc
({\it green}) and NGC 5927 ({\it red}), shifted by the differences in
distance and reddening between the clusters and M31 (see Table 1).
Instead of applying a constant shift to the entire Galactic cluster
fiducial (i.e., its ridge line and HB locus), we applied a correction
that more accurately accounts for the ACS response curve.  At each
point along the fiducial we selected a synthetic spectrum (Castelli \&
Kurucz 2003), using corresponding points on a theoretical isochrone to
determine the appropriate T$_{\rm eff}$ and log~$g$.  We then reddened
this spectrum with a Fitzpatrick (1999) extinction law, first scaled
to the cluster value and then to the M31 value.  After folding both
through the ACS bands, the difference provided the reddening vector at
that point in the fiducial. Accounting for differences in metallicity,
the agreement at the HB indicates consistency in the assumed distance
moduli, and the relative RGB positions are as expected.  However, the
subgiant branch of each Galactic cluster is significantly fainter than
that of SKHB-312.  Inspection of the CMD and the luminosity function
(LF) of each cluster shows that $\Delta m_{F814W}$ from the HB to the
subgiant branch is $\approx 0.2$~mag smaller in SKHB-312 than in
either of the Galactic clusters; this is an indication that SKHB-312
is 2--3~Gyr younger than those clusters, independent of distance,
reddening, and (to a considerable extent) metallicity assumptions.

We used 3 approaches to estimate the age of SKHB-312: fitting by eye
of the CMD to calibrated isochrones (as often done in the literature),
comparison of the CMD to population models using a Maximum Likelihood
statistic, and comparison of the LF to population models using several
different statistics.  Our models are based upon the isochrones of
VandenBerg et al.\ (2004), which agree well with CMDs of Galactic
clusters spanning a wide range of metallicity.  These isochrones have
been slightly updated from those used in Brown et al.\ (2003).  We
also used a new grid of synthetic spectra (Castelli \& Kurucz 2003) to
calculate the transformation directly from physical parameters
(T$_{eff}$, log~$g$) in the isochrones to magnitudes in the ACS bands
($m_{F606W}$, $m_{F814W}$); the updated spectra include changes to the
convective overshooting, new opacities, and the availability of
alpha-enhanced compositions.  As in our earlier work, this
transformation includes a small empirical color correction ($\lesssim
0.05$~mag) to force agreement between the transformed isochrones and
the ACS CMDs of the Galactic clusters.  This correction is smaller and
simpler than that used previously (likely due to the improvements in
the isochrones and spectral libraries), but the net result is very
similar, given the goal is agreement with the same ACS cluster
observations.  After this correction, the isochrones match the ridge
lines within $\lesssim 0.02$~mag until the saturation point in the
Galactic clusters.  When comparing isochrones to cluster data, we have
used isochrones with [$\alpha$/Fe]=0.3, and transformed them using
spectra with the same enhancement.  Note that globular clusters tend
to show alpha enhancement at all metallicities (Thomas, Maraston, \&
Bender 2003), compared to the field population.

We show the fitting by eye in Figure~1$e$, where we compare the
SKHB-312 CMD to isochrones of 8, 10, and 12~Gyr. The RGB is very
insensitive to age, but sensitive to metallicity.  Our transformed
isochrones at [Fe/H]=$-0.53$ agree well with the RGB of SKHB-312,
implying the same metallicity determined by Holland et al.\ (1997).
There is good reason to use the Holland et al.\ (1997) metallicity:
our transformation to the ACS bands cannot be verified on the upper
RGB, where the lever on metallicity is greatest, due to the saturation
of such stars in the ACS observations of Galactic clusters, while the
Holland et al.\ (1997) work was based upon WFPC2 data, which had a
very reliable transformation between ground and space bands (Holtzman
et al.\ 1995).  Clearly the main sequence turnoff and \linebreak

\vskip 0.09in

\noindent
\parbox{3.0in}{
{\sc Table 1:} M31 and Galactic cluster parameters

\begin{tabular}{lcccc}
\tableline
        & $(m-M)_V$ & $E(B-V)$ &  & age \\
Name    & (mag) &  (mag) & [Fe/H] & (Gyr)\\
\tableline
SKHB-312 & 24.68$^a$ & 0.08$^b$  & $-$0.53$^c$ & 10.0\\
47~Tuc  & 13.27$^d$  & 0.024$^e$ & $-$0.70$^f$ & 12.5\\
NGC5927 & 15.85      & 0.42      & $-$0.37$^g$ & 12.5\\
\tableline
\end{tabular}
$^a$Freedman \& Madore (1990).\\
$^b$Schlegel et al.\ (1998).\\
$^c$Holland et al.\ (1997).\\
$^d$Zoccali et al.\ (2001).\\
$^e$Gratton et al.\ (2003).\\
$^f$Kraft \& Ivans (2003).\\
$^g$Puzia et al.\ (2002).\\
}

\parbox{6.5in}{\small {\sc Fig.~1--} {\it (a)} A $24\arcsec \times
24\arcsec$ subsection of the F606W image, centered on SKHB-312, with a
logarithmic stretch.  A star near the turnoff ($m_{F814W} = 29$~mag,
$m_{F606W}-m_{F814W} = -0.4$~mag) is circled for reference. {\it (b)}
The completeness of our photometry as a function of $m_{F814W}$
(labeled) and distance from the cluster.  {\it (c)} The stellar
density as a function of distance from the cluster, for stars above
the turnoff ($m_{F814W} \le 29$), normalized to unity ({\it dashed})
beyond the tidal radius of the cluster, uncorrected for
incompleteness.  Our CMD analysis is restricted to the annulus
spanning 100 pix -- 300 pix ({\it shaded}), i.e., from the point where
the star counts roll over due to severe crowding, to the point where
the field contamination is 50\%.  {\it (d)} The CMD from the annulus
shown in the previous panel.  Note the tight clump of red HB stars
(unobscured in the next panel) and the red RGB locus, compared to the
CMD of the field population (Brown et al.\ 2003).  The turnoff is
clearly detected.  The ridge lines ({\it curves}) and HB locus ({\it
points}) from two Galactic clusters ({\it labeled}) are overplotted.
The clusters bracket the metallicity of SKHB-312, yet the turnoff of
each is fainter than that of SKHB-312. The luminosity difference
between the HB and the subgiant branch is $\approx$0.2~mag smaller in
SKHB-312 than in the Galactic clusters, indicating an age 2--3 Gyr
younger.  {\it (e)} The CMD of SKHB-312 with a 10~Gyr isochrone ({\it
yellow}) at the cluster metallicity shows good agreement; 8~Gyr ({\it
green}) is clearly too young, and 12~Gyr ({\it red}) is too old. {\it
(f)} A 10~Gyr isochrone plotted with the same halo contamination,
scattering, and completeness ({\it shaded}) as in the data; beyond the
CMD region used for fitting, the SKHB-312 data are shown.  This 10~Gyr
model agrees well with the same region shown in the previous panels.}

\vskip 0.1in
\epsfxsize=6.5in \epsfbox{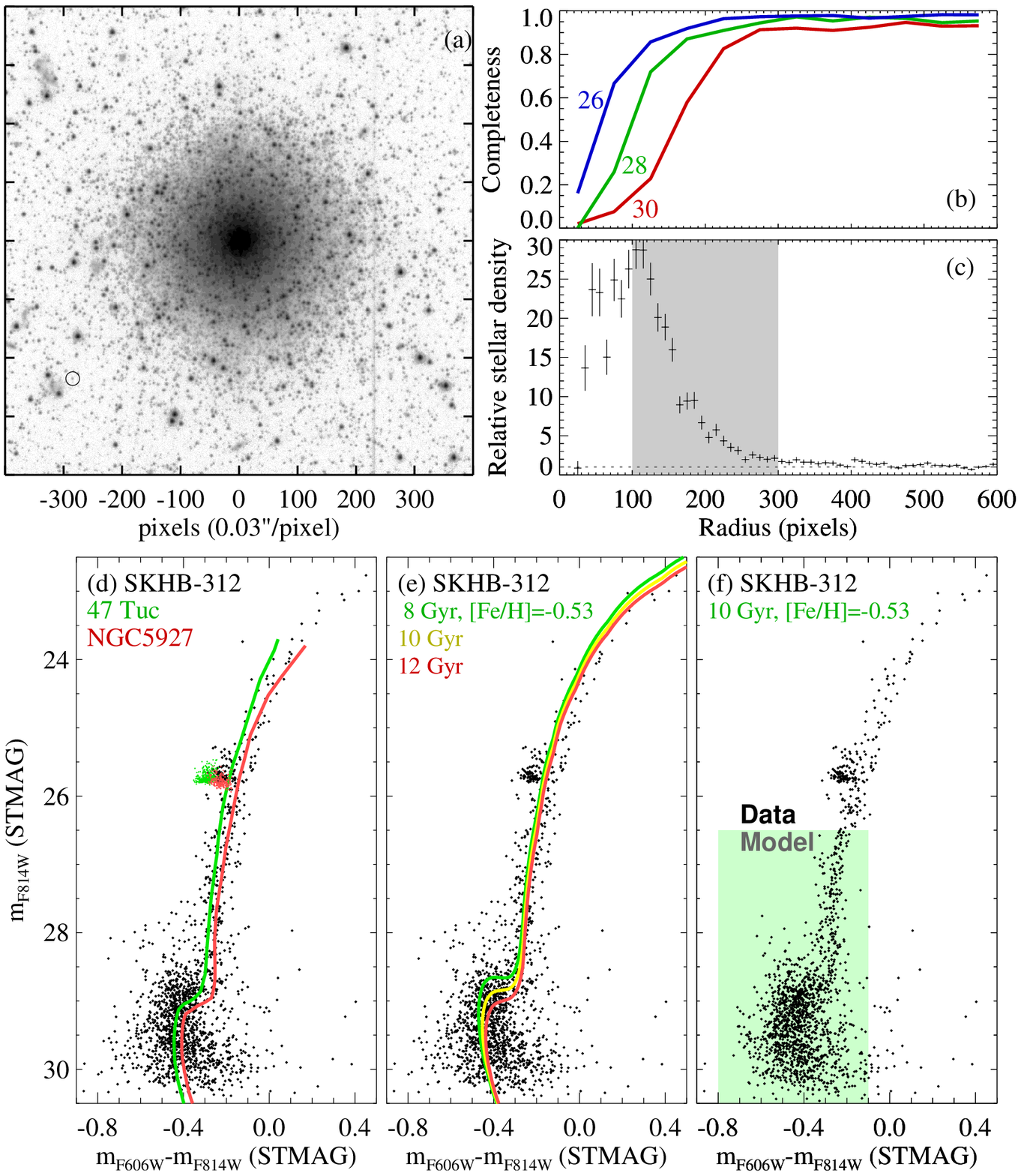} 

\newpage

\noindent
subgiant branch in Figure~1$e$ are too bright at 8~Gyr, and too faint
at 12~Gyr, implying an age of $\approx$10$\pm 1$~Gyr for SKHB-312.
However, fitting by eye ignores two subtle effects that might bias the
fit toward younger ages: contamination by field stars, which are known
to span a wide age range (Brown et al.\ 2003), and blends of stars in
the crowded photometry, which can brighten the main sequence turnoff
and subgiant branch.  These effects are included in our quantitative
modeling below.

We compared CMD models to the data using a Maximum Likelihood
statistic: $\sum $~ln~$P_i$, where $P_i$ is the probability of a star
falling at its observed CMD location, given the model under
consideration.  Each model gives the expected distribution of stars in
the CMD, on a 0.02~mag grid, so that the probability of a star being
observed in a given bin ($P_i$) is the fraction of the model falling
in that bin (see, e.g., Dolphin 2002). The probability that all of the
stars were observed in their given CMD bins is the product of the
individual probabilities, and the Maximum Likelihood statistic is a
logarithmic score reflecting the probability of observing the SKHB-312
CMD under a given model.  We focused on the CMD region highlighted in
Figure~1$f$: $-0.8 \le (m_{F606W}-m_{M814W}) \le -0.1$~mag and $26.5
\le m_{F814W} \le 30.5$~mag, as in Brown et al.\ (2003), which
includes 1456 of the 1720 stars in the SKHB-312 catalog.

To construct each synthetic CMD, we first populated our isochrones
with a Salpeter initial mass function (IMF), and then applied the
photometric errors, blends, and incompleteness calculated in the
artificial star tests, using the Starfish code (Harris \& Zaritsky
2001).  We then modeled the halo field contamination by applying the
photometric errors, blends, and incompleteness of the SKHB-312 annulus
to the best-fit combination of isochrones in Brown et al.\ (2003),
which reproduces the halo data very well.  Note that one cannot simply
statistically subtract the field data from the cluster data, because
they have dramatically different noise and completeness.  After
normalizing the field model to the area of the SKHB-312 annulus, we
found that the halo field contamination in the SKHB-312 CMD was 8\% by
number. We combined the cluster model and field contamination model to
create a synthetic CMD for comparison to the data, with the total
number of stars normalized to 1456. This was repeated for isochrone
ages of 5--15~Gyr with 0.5~Gyr steps, with the likelihood statistic
calculated for each, giving a best-fit of 11~Gyr.  To evaluate the
goodness of fit, we created Monte Carlo realizations of the best-fit
model and evaluated the likelihood statistic.  The likelihood values
for these samples were higher than for the model tested against the
real data, suggesting that the models are imperfect (e.g., because of
binaries, blue stragglers, a non-Salpeter IMF, or remaining
imperfections in the isochrones).  We reproduced the likelihood values
seen for the best fit by distributing 60 stars (4\% of the sample)
randomly within the CMD region being fit.  We used the distribution of
the Monte Carlo scores (including 60 randomly distributed stars) to
assign confidence intervals to the ages measured from the real data.
The results imply age constraints of $9.5 \le$~age~$\le 12.5$~Gyr at
68.3\% confidence and $9.0 \le$~age~$\le 14.0$~Gyr at 99.7\%
confidence.  Thus, the likelihood fitting of the CMD is consistent
with our ``by eye'' age estimate of 10~Gyr.  Figure~1$f$ shows the CMD
of SKHB-312 outside of the fitting region, and a realization of the
10~Gyr model within the fitting region.  Comparison of the model to
the data (Figure~1$e$) shows good agreement.

Finally we fit the SKHB-312 LF, which is often preferred when fitting
data with few stars and/or large photometric errors.  We collapsed the
model CMDs and the data CMD in color, and then compared them using
three statistics: $\chi^2$, $\chi^2_\gamma$ (Mighell 1999), and the
likelihood statistic of Dolphin (2002).  The best fit in each case,
respectively, is 10~Gyr (9--12.5~Gyr at 68.3\% confidence), 10.5~Gyr
(9--13.0~Gyr at 68.3\% confidence), and 10~Gyr (8.5--12.5~Gyr at
68.3\% confidence); the confidence intervals come from Monte Carlo
tests, comparing the best model against realizations of that model
(artificial data), as done in the Maximum Likelihood fitting above.
These results are also consistent with the likelihood fitting to the
CMD and our fitting by eye, implying an age of $\approx$10~Gyr for
SKHB-312.

\section{Summary}

We have directly determined the age of the M31 cluster SKHB-312, by
fitting its deep photometry extending below the main sequence turnoff.
These data presented a difficult challenge because the cluster is so
compact, and despite our extensive efforts, we do not find tight
formal constraints on its age: 10$_{-1}^{+2.5}$~Gyr (or $\approx
9$~Gyr if core He diffusion were included in the models).  However,
SKHB-312 appears to be 2--3~Gyr younger than Galactic clusters at
similar metallicities. Although these data prove that direct ages can
be determined for M31 clusters with existing instrumentation, a
program seeking tighter constraints would benefit from the use of the
ACS High Resolution Camera and the choice of a less compact cluster.

\acknowledgements

Support for proposal 9453 was provided by NASA through a grant from
STScI, which is operated by AURA, Inc., under NASA contract NAS
5-26555.  We are grateful to J.\ Harris and P.\ Stetson for providing
their codes and assistance.

\end{document}